\documentclass[aip,reprint,superscriptaddress]{revtex4-1}
\usepackage{graphicx}% Include figure files
\usepackage{epstopdf} % Fixes compile error that file figure is not found.
\usepackage{bm}% bold math
\usepackage{dcolumn}% Align table columns on decimal point
\usepackage{amssymb}
\usepackage{amsmath}
\usepackage{xcolor}

\begin{document}

\title{Ultrathin perpendicular free layers for lowering the switching current in STT-MRAM}

\author{Tiffany S.\ Santos}
	\email{tiffany.santos@wdc.com}
	\affiliation{Western Digital Research Center, Western Digital Corporation, San Jose, CA 95119\\}
\author{Goran Mihajlovi\'{c}}
\author{Neil Smith}
\author{J.-L.\ Li}
\author{Matthew Carey}
\author{Jordan A.\ Katine}
\author{Bruce D.\ Terris}
	
\date{\today}

\begin{abstract}
The critical current density $J_{c0}$ required for switching the magnetization of the free layer (FL) in a spin-transfer torque magnetic random access memory (STT-MRAM) cell is proportional to the product of the damping parameter, saturation magnetization  and thickness of the free layer, $\alpha M_S t_F$. Conventional FLs have the structure CoFeB/nonmagnetic spacer/CoFeB. By reducing the spacer thickness, W in our case, and also splitting the single W layer into two layers of sub-monolayer thickness, we have reduced $t_F$ while minimizing $\alpha$ and maximizing $M_S$, ultimately leading to lower $J_{c0}$ while maintaining high thermal stability. Bottom-pinned MRAM cells with device diameter in the range of 55--130~nm were fabricated, and $J_{c0}$ is lowest for the thinnest (1.2~nm) FLs, down to 4~MA/cm$^2$ for 65~nm devices, $\sim$30\% lower than 1.7~nm FLs. The thermal stability factor $\Delta_{\mathrm{dw}}$, as high as 150 for the smallest device size, was determined using a domain wall reversal model from field switching probability measurements. With high $\Delta_{\mathrm{dw}}$ and lowest $J_{c0}$, the thinnest FLs have the highest spin-transfer torque efficiency.
\end{abstract}
\maketitle

\section{Introduction}

A leading challenge for the realization of a high-density STT-MRAM product is to have a low critical current required for switching the magnetization of the free layer, while maintaining high thermal stability for data retention. One route to reducing $J_{c0}$ is to lower $M_S t_F$.\cite{Sun_PRB2000} However, reducing $M_S$ is known to increase the temperature dependence of the magnetic anisotropy and thermal stability factor,\cite{Iwata-Harms_SR2018} which leads to poor retention at elevated operating temperature. Rather than reducing $M_S$, the approach taken in this study is to reduce $t_F$ while maximizing $M_S$. Conventional free layers have the structure CoFeB/nonmagnetic spacer/CoFeB with a total thickness close to 2~nm. The role of the nonmagnetic spacer\textemdash typically metals such as Ta,W or Mo\textemdash is to absorb the boron atoms during the anneal step needed for solid-phase epitaxy of the MgO barrier and the CoFeB electrodes, which is critical for achieving high tunnel magnetoresistance (TMR)\cite{Butler_PRB2001, Mathon_PRB2001} and high interfacial perpendicular anisotropy.\cite{Sato_APL2012, Kim_SR2015, Naik_AIPAdv2012} The drawbacks for inserting these boron-sink materials are that they increase the damping constant of the FL, which is disadvantageous for lowering $J_{c0}$, and they create regions of low or zero moment in the free layer, referred to as a dead layer, which reduce volume-averaged $M_S$.

In this study the strategy for increasing $M_S$ was to reduce the spacer thickness (and thus the dead layer) as much as possible, while still maintaining perpendicular anisotropy. Thicker CoFeB films need thicker spacer layers to stabilize perpendicular anisotropy. Hence, continuous reduction in spacer thickness necessitated a reduction in FL thickness. We found that thinner FLs with thinner spacer layers have higher $M_S$.  Our film characterization shows that FL properties improved by reducing $t_F$ down to 1.2~nm and splitting the single spacer layer into two thinner spacer layers. Our ultrathin FLs have $M_S$ reaching above 1700~emu/cm$^{3}$ with no dead layer and with $\alpha$ of only 0.003. We tested devices with $t_F$ in the range of 1.2\textendash1.7~nm and device size down to $\sim$55~nm.  Our device testing shows that $J_{c0}$ is lower for the thinner FLs while the thermal stability factor for domain wall reversal $\Delta_{\mathrm{dw}}$ remains high, so that overall there is an increase in spin-transfer torque efficiency, defined as $\Delta_{\mathrm{dw}}/I_{c0}$ where $I_{c0}$ is the critical switching current.

\section{Experimental Method}

The STT-MRAM stacks in this study consist of the following layer structure: bottom electrode/seed layer/CoFeB reference layer/MgO tunnel barrier/free layer/MgO cap/Ru/Ta/Ru top electrode. In these bottom-pinned stacks, a Co/Pt-based synthetic antiferromagnet (SAF) was used to pin the magnetization of the reference layer. These stacks were sputtered at room temperature using an Anelva-7100 system and then postannealed at 335$^\circ$C for 1 hour. The MgO layers were RF-sputtered from a MgO target.

In order to study the magnetic properties of the FL without any additional magnetic signal from the SAF or reference layer in the full STT-MRAM stack, the following FL-only stacks were prepared on oxidized Si substrates (thickness in nm): bottom electrode/3 Ta/5 Ru/0.3 W/0.5 CoFeB/1 MgO/FL/0.7 MgO/Ru. The 0.5~nm CoFeB seed layer is nonmagnetic at room temperature. The magnetic properties of the FL in the FL-only stack closely match those of the identical FL deposited in the bottom-pinned MRAM stack.

A FL thickness ladder of the following thicknesses were prepared: 1.2, 1.3, and 1.7~nm. For each thickness, two FL designs were compared: one with a single tungsten spacer (referred to as single-W) and the other with two tungsten spacers (referred to as split-W). For a given FL thickness, the combined W thickness of the two W layers in the split-W FL is equivalent to the thickness of the single W spacer in the single-W FL. The total W thickness is designated as $t_W$. The single-W FLs have the following layer structure, listed in order as sputtered on top of the MgO barrier: CoFeB/CoFe/Mg/W/CoFe. The split-W FLs have the following layer structure: CoFeB/CoFe/Mg/W/CoFe/Mg/W/CoFe. Refer to Table~\ref{layer_table} for the nominal thickness of every layer for each FL in this study. The Mg layers are sacrificial layers that protect the previously sputtered CoFe film from damage caused by the impinging W atoms during deposition of the W spacer, as described in Ref.~\onlinecite{Swerts_APL2015}. In agreement with Ref.~\onlinecite{Swerts_APL2015}, we have found that no Mg remains in the FL.\footnote{As determined in a separate study of similar film stacks using cross-section transmission electron microscopy with elemental analysis using electron energy loss spectroscopy, in both the as-sputtered and post-annealed state} The Mg thickness used here was determined by optimizing for highest $M_S H_k$ of the free layer. $M_S$ and anisotropy field $H_k$ were determined from hard-axis magnetization versus magnetic field $M(H)$ loops of unpatterned films measured by vibrating sample magnetometry (VSM).

\begin{table*}
\caption{\label{layer_table} Nominal thickness (in nm) of each layer within the FLs compared in this study. The single-W free layers have the structure: CoFeB/CoFe/Mg/W/CoFe, and the split-W free layers have the structure: CoFeB/CoFe/Mg/W/CoFe/Mg/W/CoFe.}
	\begin{ruledtabular}
		\begin{tabular}{ccccccccccccccc}
		& & \multicolumn{5}{c}{single-W} & \multicolumn{8}{c}{split-W} \\
		\cline{3-7} \cline{8-15} 
			$t_F$ &$t_W$ &CoFeB &CoFe &Mg &W &CoFe &CoFeB &CoFe &Mg &W &CoFe &Mg &W &CoFe\\
		\hline
			1.22 &0.04 &0.47 &0.31 &1.00 &0.04 &0.40 &0.47 &0.11 &0.60 &0.02 &0.26 &0.60 &0.02 &0.34\\
			1.20 &0.015 &0.47 &0.31 &1.00 &0.015 &0.40 &0.47 &0.11 &0.60 &0.008 &0.26 &0.60 &0.007 &0.34\\
			1.30 &0.04 &0.52 &0.32 &1.00 &0.04 &0.42 &0.52 &0.10 &0.60 &0.02 &0.32 &0.60 &0.02 &0.32\\
			1.70 &0.12 &0.73 &0.32 &1.00 &0.12 &0.53 &0.73 &0.11 &0.60 &0.06 &0.37 &0.60 &0.06 &0.37\\
			
\end{tabular}
\end{ruledtabular}
\end{table*}

\begin{figure}
\includegraphics{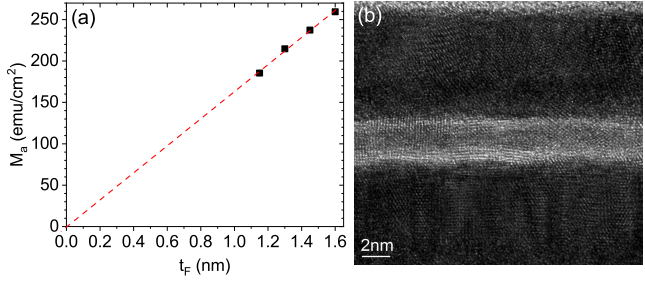}
\caption{\label{deadlayer_XSTEM} a) Saturation moment divided by sample area $M_a$ versus $t_F$ for a FL-only thickness ladder with the structure CoFeB/CoFe/0.6~Mg/0.02~W/CoFe/0.6~Mg/0.02~W/CoFe, in which only the CoFeB and CoFe thicknesses are varied. The linear fit intersects the x-axis at 0~nm, indicating that there is no dead layer. b) Cross-section TEM of the $t_F$=1.3~nm sample with moment shown in (a) and structure given in Table~\ref{layer_table}.}
\end{figure}

The $t_F$ reported here is the nominal film thickness and includes the thickness of all the CoFeB, CoFe and W components. The nominal film thickness was determined using sputter time, and the sputter rate (i.e. nm/s) was calibrated for thick films using x-ray reflectivity. Thus, as the nominal W layer thickness is well below the thickness of a W atom, the W layer thickness refers to a partial monolayer of W atoms. As shown in the plot of saturation moment per area versus $t_F$ in Fig.~\ref{deadlayer_XSTEM}a for our split-W free layer design with $t_W$=0.04~nm, the linear fit line intersects the x-axis at 0~nm, indicating that there is no dead layer. A cross-section transmission electron micrograph (TEM) of the split-W FL-only film with $t_F$=1.3~nm is shown in  Fig.~\ref{deadlayer_XSTEM}b. The 1.0~nm thick MgO barrier and the 0.7~nm MgO cap are easily identified, with the 1.3~nm free layer in between the MgO layers.

\begin{figure}
\includegraphics{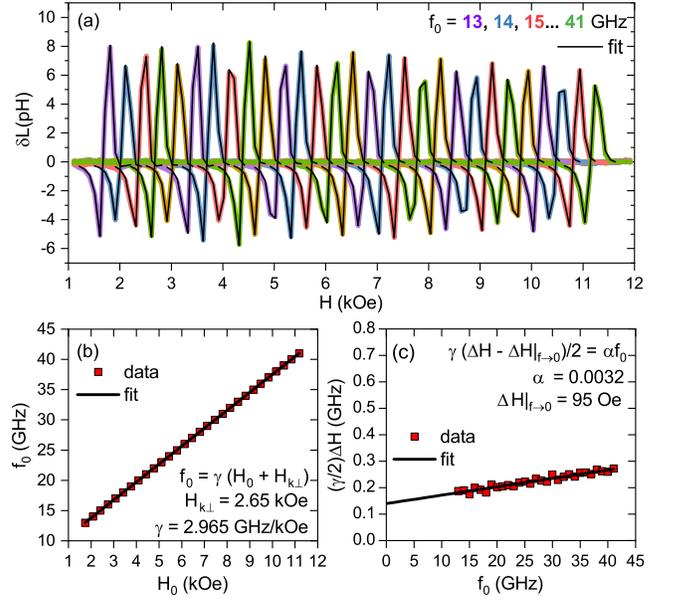}
\caption{\label{FMRplot} FMR data for the split-W FL with $t_F$=1.3~nm. (a) From VNA-measured $S_{21}(H;f_0)$, extracted $\delta L(H;f_0)=-(50\Omega/\pi f_0) \delta \ln|S_{21}|$, the magnetic film contribution to lossy component of inductance of coplanar waveguide and sample, where ``$\delta$'' imples a \textit{finite} difference, e.g., $\delta L(H;f_0)=L(H+\delta H;f_0)-L(H-\delta H;f_0)$. Here, $\delta H$=100~Oe. The black lines are least squares fits: $\delta L(H;f_0) \propto$ Im $\delta \chi (H) \cos{\phi} (f_0) - $ Re $\delta \chi (H) \sin{\phi} (f_0)$, where $\chi (H) \propto (H-H_0 - i \Delta H/2)^{-1}$ with fit parameters $H_0 (f_0)$, linewidth $\Delta H(f_0)$, and phase $\phi$ to account for other loss factors.\cite{Smith_unpublished} (b) Linear fit: $f_0 = \gamma (H_0 + H_{k\perp})$ to obtain $\gamma$ and $H_{k\perp}$. (c) Linear fit: $\gamma (\Delta H - \Delta H|_{f\rightarrow0})/2=\alpha f_0$ to obtain $\Delta H|_{f\rightarrow0}$ and ${\alpha}$.}
\end{figure}

Vector network analyzer ferromagnetic resonance (VNA-FMR) measurements were performed on the unpatterned, FL-only films to determine the effective anisotropy field $H_{k\perp}$, the damping parameter $\alpha$, and inhomogeneous linewidth broadening $\Delta H|_{f\rightarrow0}$. The applied field $H$ perpendicular to the film plane was swept once from 12~kOe to 1~kOe while transmission coefficients $S_{21}(H;f)$ were simultaneously measured at 29 discrete frequencies in the range $f_0$=13--41~GHz. See Fig.~\ref{FMRplot} for an example. Inhomogeneous broadening originates from non-uniformities of the anisotropy, and thus it is an indicator of film uniformity. Lower $\Delta H|_{f\rightarrow0}$ indicates better film uniformity.

Device performance was compared for the FLs in the thickness ladder, specifically the six free layers described in rows 2-4 of Table~\ref{layer_table}. Bottom-pinned STT-MRAM test devices were fabricated using 193~nm deep UV optical lithography, followed by reactive ion etching of a hard mask, ion milling of the MRAM film, SiO$_2$ refill and then chemical mechanical planarization. Four different device sizes were fabricated for each FL, resulting in median electrical diameters $D$ in the range of 55 to 130~nm. Device diameter was determined from the measured device resistance in the parallel state $R_P$ and the resistance-area product $RA$, as $D=\sqrt{4RA/\pi R_P)}$. The $RA$ product, along with TMR ratio, was measured using the current-in-plane-tunneling (CIPT) technique\cite{Worledge&Trouilloud_APL2003} on identical, unpatterned film stacks sputtered within the same batch as the device wafers.

\begin{figure}
\includegraphics[width=8.5cm]{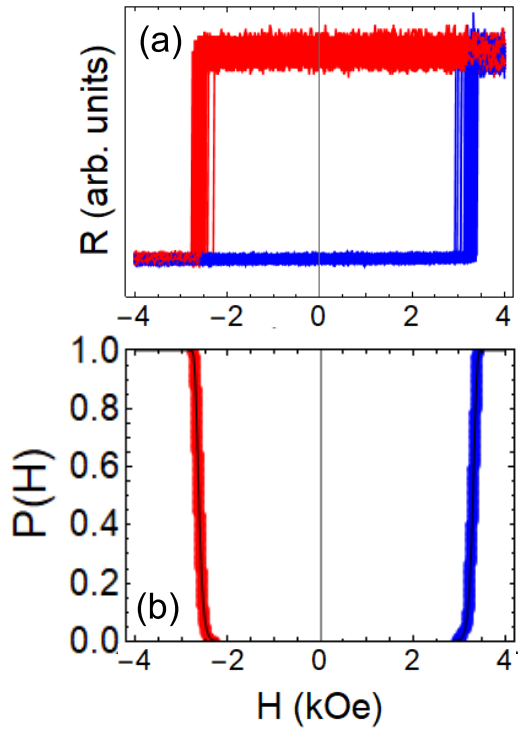}
\caption{\label{fieldswitchingplots} (a) Resistance $R$ versus field loops for a 63~nm device with a split-W FL having $t_F$=1.3~nm. (b) Empirical $P(H)$ corresponding to the switching fields in (a). The black line is the fit to a domain wall reversal model,\cite{Mihajlovic_inpreparation} yielding $\varepsilon_{\mathrm{dw}}$=8.4~erg/cm$^2$, $w_{\mathrm{dw}}$=12.5~nm and $\Delta_{\mathrm{dw}}$=164.1 for this device.}
\end{figure}

To determine $\Delta_{\mathrm{dw}}$ for the FLs in the devices, we measured empirical field-switching probability distributions $P(H)$, whereby 100 resistance versus field transfer loops were measured at low bias (10mV, to eliminate spin-transfer torque effects) using a staircase ramp with a field step of 5~Oe and a dwell time of 0.2~ms. An example is shown in Fig.~\ref{fieldswitchingplots}a for a device with diameter 63~nm and split-W FL with $t_F$=1.3~nm. Because the reversal mechanism for the FL magnetization in devices with diameters larger than 50~nm that we have studied here is by domain wall nucleation and propagation,\cite{Thomas_IEDM2015, Chaves-OFlynn_PRA2015} $P(H)$ was fit using a domain wall reversal model.\cite{Mihajlovic_inpreparation} An example of the fit is shown as black lines in Fig.~\ref{fieldswitchingplots}b. The fit parameters are domain wall energy density $\varepsilon_{\mathrm{dw}}$ and domain wall width $w_{\mathrm{dw}}$, and $\Delta_{\mathrm{dw}}$ is given as $\Delta_{\mathrm{dw}}=D \varepsilon_{\mathrm{dw}} t_F / k_B T$, where $k_B$ is the Boltzmann constant and $T$ is temperature. Using the following relations between $\varepsilon_{\mathrm{dw}}$ and $w_{\mathrm{dw}}$, we calculate exchange constant $A_{\mathrm{ex}}$:

\begin{eqnarray}
\varepsilon_{\mathrm{dw}}=4 \sqrt{\frac{1}{2} M_S H_{k} A_{\mathrm{ex}}}
\\
w_{\mathrm{dw}} = 2 \ln{2} \sqrt{\frac{2 A_{\mathrm{ex}}}{M_S H_{k}}}
\\
A_{\mathrm{ex}} = \frac{w_{\mathrm{dw}} \varepsilon_{\mathrm{dw}}}{8 \ln{2}}
\end{eqnarray}

The $J_{c0}$ was obtained by fitting the pulse width ($t_P$) dependence of switching voltage $V_c$, measured in the range of 5~$\mu$s - 5~ms as described in Ref.~\onlinecite{Mihajlovic_PRA2020}, to the thermal activation model for spin-transfer torque switching.\cite{Koch_PRL2004, Li_PRB2004} The switching voltage $V_{c0}$ (defined at zero temperature and infinitely long time) is the intercept of the linear fit for $V_c$ versus $\ln{t_P}$, and $J_{c0}=V_{c0}/RA$.

\section{Results and Discussion}
\subsection{Film properties}

\begin{figure*}
\includegraphics{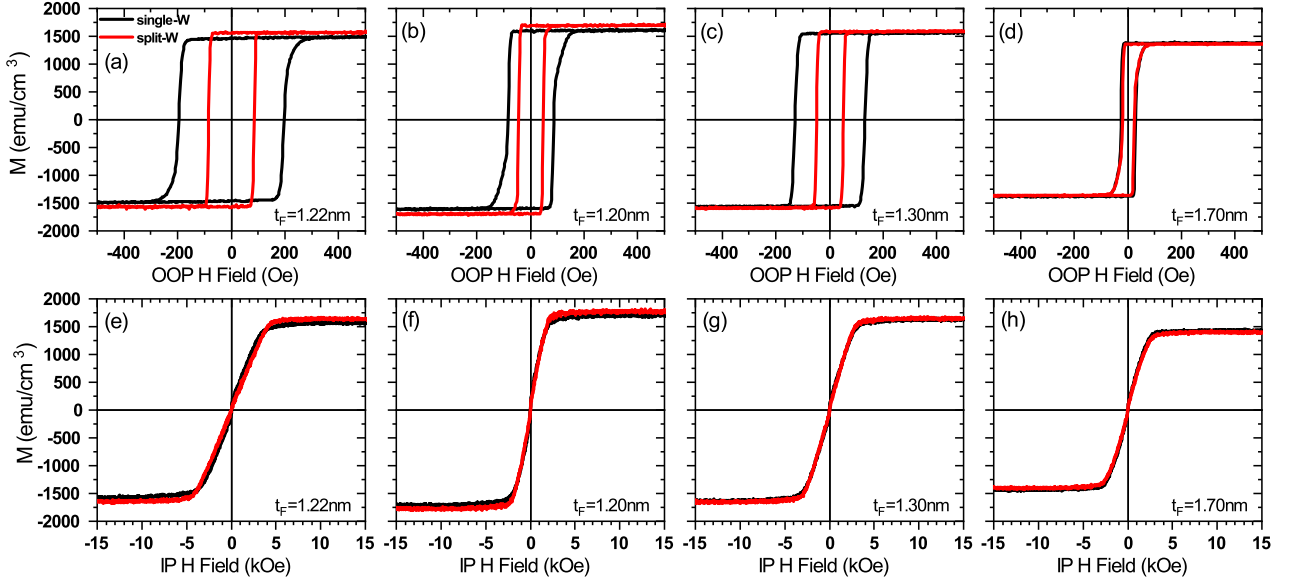}

\caption{\label{MHall} $M(H)$ loops for FL-only films with the indicated $t_F$, for $H$ applied out-of-plane (OOP) along the easy axis (a,b,c,d) and for $H$ applied in-plane (IP) along the hard axis (e,f,g,h). Description of all samples is in Table~\ref{layer_table}, and the $M_S$ values are listed in Table~\ref{FMRchart}.}
\end{figure*}

We first compare the magnetic properties for a single-W FL and a split-W FL with the same nominal $t_F = 1.22$~nm and $t_W = 0.04$~nm. Figures \ref{MHall}a and \ref{MHall}e display the easy axis and hard axis $M(H)$ loops, respectively, for the FLs described in the first row of Table~\ref{layer_table}. The split-W FL has a slightly higher $M_S$ and sharper switching (more square loop), as seen in the easy axis $M(H)$, and higher anistropy field, as seen in the hard axis $M(H)$. The $M_S$ and FMR results are listed in Table~\ref{FMRchart}. Splitting the W spacer from a single 0.04~nm layer to two 0.02~nm layers resulted in an sizeable increase of $H_{k\perp}$ from 3.02~kOe to 4.07~kOe and more than 3-fold reduction of $\Delta H|_{f\rightarrow0}$, signifying a more uniform film. The low damping constant was maintained at $\alpha=0.004$. 

It has been our observation that $\Delta H|_{f\rightarrow0}$ measured at the film level has a negative correlation to the coercive field $H_C$ at the device level; higher $\Delta H|_{f\rightarrow0}$ typically results in low $H_C$, regardless of $H_k$. Therefore, it is desireable to minimize $\Delta H|_{f\rightarrow0}$ as much as possible. In order to lower $\Delta H|_{f\rightarrow0}$ even further, we reduced the thickness of the W spacer. Figures \ref{MHall}b and \ref{MHall}f display the M(H) loops for the single-W and split-W FLs for which $t_W$ was reduced to 0.015~nm (refer to the second row in Table~\ref{layer_table}). As is typical when reducing the W spacer thickness, we observed higher $M_S$, lower $H_{k\perp}$ and lower $\alpha$, compared to the $t_W=0.04$~nm FLs. Importantly, there was another significant reduction of $\Delta H|_{f\rightarrow0}$ when going from single- to split-W spacers, dropping to $<$100~Oe which is promising for high $H_C$ at the device level.

\begin{table*}
\caption{\label{FMRchart} Materials parameters for all FLs in this study: $M_S$ values measured by VSM, FMR results ($H_{k\perp}$, $\alpha$, and $\Delta H|_{f\rightarrow0}$), and $A_{\mathrm{ex}}$ determined from devices using Equation~3. The parameter values are expressed to the last significant digit based on the corresponding error analysis, except for $\alpha$ where the absolute error is 0.0003.}
\begin{ruledtabular}
		\begin{tabular}{cccccccc}
			$t_F$ &$t_W$ &No. of &$M_S$  &$H_{k\perp}$  &$\alpha$ &$\Delta H|_{f\rightarrow0}$ &$A_{\mathrm{ex}}$\\
			(nm) &(nm) &W layers &(emu/cm$^3$)& (kOe)& &(Oe) &($\mu$erg/cm)\\
		\hline
			1.22 &0.04 &1 &1572 &3.02 &0.0040 &420 \\
			1.22 &0.04 &2 &1644 &4.07 &0.0037 &125\\
			1.20 &0.015 &1 &1714 &1.44 &0.0036 &210 &1.7\\
			1.20 &0.015 &2 &1783 &1.72 &0.0031 &80 &2.01\\
			1.30 &0.04 &1 &1634 &2.55 &0.0037 &180 &1.77\\
			1.30 &0.04 &2 &1651 &2.65 &0.0032 &95 &1.85\\
			1.70 &0.12 &1 &1435 &1.98 &0.0044 &95 &1.49\\
			1.70 &0.12 &2 &1400 &2.05 &0.0044 &85 &1.47\\			
\end{tabular} 
\end{ruledtabular}
\end{table*}

For the $t_F$=1.3~nm and 1.7~nm films in the FL thickness ladder of this study, the $M(H)$ loops are displayed in Figures~\ref{MHall}c,g and \ref{MHall}d,h, respectively, and the magnetic properties are listed in Table~\ref{FMRchart}. Thicker FLs require more W in order to maintain perpendicular anisotropy, which results in lower $M_S$. Therefore, $M_S$ is highest for the thinnest FLs. As demonstrated above, $H_{k\perp}$ can be tuned by the W amount. The W amount used here is the minimum amount necessary to maintain perpendicular anisotropy. $\Delta H|_{f\rightarrow0}$ is significantly lower for the split-W FLs compared to the single-W FLs, with the exception of the 1.7~nm FLs, for which there appears to be no advantage to having two spacers as all parameters are nearly the same for the two designs.  Noteably, contrary to the common observation that $\alpha$ and $\Delta H|_{f\rightarrow0}$ increase for thin CoFeB,\cite{Liu_JAP2011,Sabino_IEEETransMag14,Enobio_IEEEMagLett2015} $\alpha$ is very low for all of these ultrathin FLs, which is advantageous for achieving lower switching current.

\subsection{Device Performance}
\begin{figure}
\includegraphics{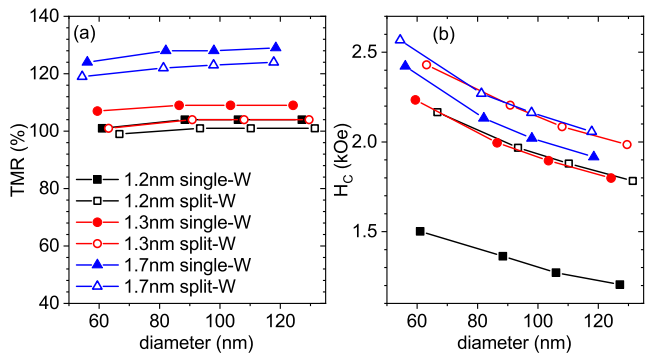}
\caption{\label{TMRandHc} Device TMR and coercivity $H_C$ versus electrical diameter. Standard error of the mean is smaller than the data points.}
\end{figure}

The TMR for the six FLs is shown versus electrical diameter in Fig.~\ref{TMRandHc}a. This TMR measured for the devices closely matches the TMR values measured by CIPT (not shown). TMR is lower for the thinner FLs. In addition, TMR is lower for the split-W FLs. In general, lower TMR is observed for thinner free layers.\cite{Sato_IEEEMagnLett2012, Cuchet_APL2014} More specifically in our FLs, in the single-W FL stack as $t_F$ becomes thinner, the amount of CoFeB and CoFe between the MgO barrier and the W spacer becomes thinner, which means that the W atoms are closer to the MgO barrier. Any W atoms near the MgO/CoFeB interface are expected to disrupt the solid-phase epitaxy across this interface and thus disturb the coherent tunneling that otherwise generates high TMR.\cite{Butler_PRB2001} The W atoms are even closer to the MgO/CoFe interface in the case of the split-W FLs, which is thought to cause the TMR to be even lower.

The device coercivity $H_C$ (see Fig.~\ref{TMRandHc}b) is in the range of $\sim$2.2--2.6~kOe for the smallest device size for all FL designs, with the exception of the 1.2~nm single-W FL, which has the lowest film-level $H_{k\perp}$ and highest $\Delta H|_{f\rightarrow0}$.

\begin{figure}
\includegraphics{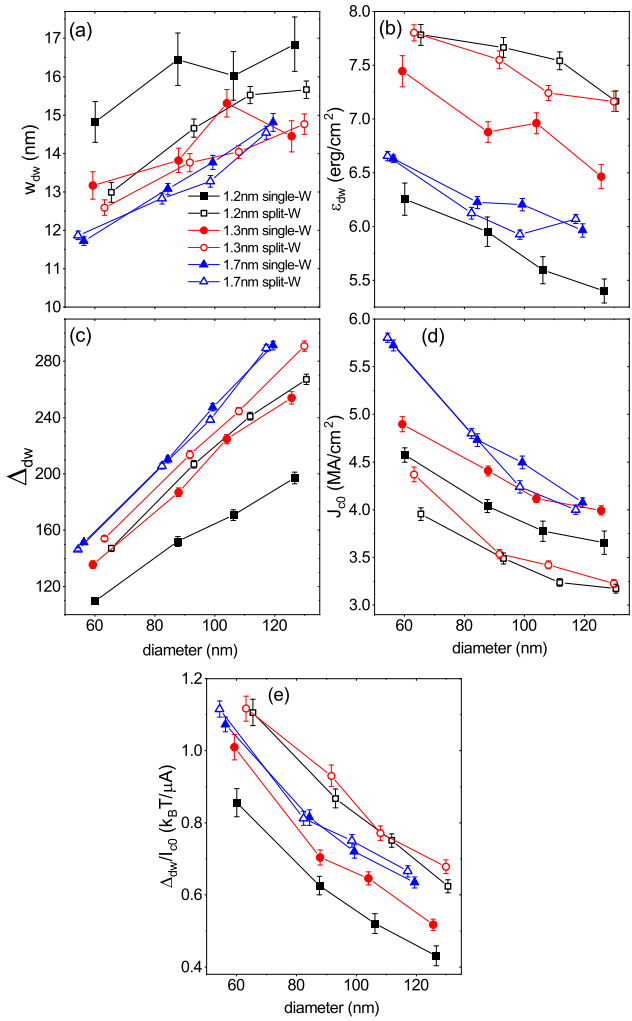}
\caption{\label{device_data} (a) $w_{\mathrm{dw}}$, (b) $\varepsilon_{\mathrm{dw}}$, (c) $\Delta_{\mathrm{dw}}$, (d) $J_{c0}$ and (e) $\Delta_{\mathrm{dw}}/I_{c0}$ versus electrical diameter for all studied FLs. Each data point is the median value of approximately 35 measured devices, and the error bar is the standard error of the mean.}
\end{figure}

The results of the field switching probability test with data fit to a domain wall reversal model is displayed in Figure~\ref{device_data}a-c. The $w_{\mathrm{dw}}$ for all FLs is the range of 12--17~nm. The 1.2~nm and 1.3~nm split-W FLs have the highest $\varepsilon_{\mathrm{dw}}$. When multiplied by free layer volume to calculate $\Delta_{\mathrm{dw}}$, we find that all FLs, with the exception of the single-W 1.2~nm FL (having low $H_C$), have similarly high $\Delta_{\mathrm{dw}}$ in the range of 130--150 for the smallest device size. Noteably, for the split-W FLs, we have reduced $t_F$ from 1.7~nm to 1.2~nm without significant loss of $\Delta_{\mathrm{dw}}$, while $J_{c0}$ drops by a third with this $t_F$ reduction, as shown in Fig.~\ref{device_data}d. Indeed, $J_{c0}$ is lowest for the thinnest FLs with the split-W design, which have the lowest $\alpha$ (refer to Table~\ref{FMRchart}). Spin-transfer torque efficiency, defined as $\Delta_{\mathrm{dw}}/I_{c0}$ and shown in Fig.~\ref{device_data}e, is highest for the thinnest FLs with the split-W design.

The $A_{\mathrm{ex}}$ for each FL calculated using Equation~3 is listed in Table~\ref{FMRchart}. Overall, these $A_{\mathrm{ex}}$ values are high, approaching the values for bulk Fe (2.0~$\mu$erg/cm) and confirming the high quality of the FLs, in agreement with the film-level characterization. These values are similar to $A_{\mathrm{ex}}$ deduced from spin-wave spectroscopy for perpendicular CoFeB free layers in nanopillar magnetic tunnel junctions.\cite{Helmer_PRB2010, Devolder_JAP2016} For the 1.2~nm and 1.3~nm FL thicknesses, $A_{\mathrm{ex}}$ is higher for the split-W FL compared to the single-W FL, consistent with the superior properties observed at the film level for the split-W design (refer to Table~\ref{FMRchart}). Whereas for the two 1.7~nm FLs, the $A_{\mathrm{ex}}$ values are nearly the same, consistent with the other parameters for these films listed in Table~\ref{FMRchart}. When comparing $A_{\mathrm{ex}}$ across the thickness ladder for the split-W FLs, we observe higher $A_{\mathrm{ex}}$ for thinner W spacer, in agreement with the trend observed with the film-level magnetometry measurements of Ref.~\onlinecite{Mohammadi2019} and magneto-optical Kerr microscopy studies of Ref.~\onlinecite{Buford_IEEEML2016a}.

As the film and device results for the 1.7~nm single-W and split-W FLs are the same, one may conclude that there is no advantage to splitting the single W spacer into two spacers for this FL thickness, and that this split-W approach is only beneficial for thinner FLs. However, only one 1.7~nm split-W FL was explored in this study. By varying the thickness of the CoFeB and CoFe layers below, between, and above the two W spacers, it may be possible to tune $M_S$ and $H_k$ to achieve a 1.7~nm split-W FL design with superior performance. This would suggest some sensitivity to the proximity of the two W spacers to the MgO interfaces and to each other. Of course, thicker CoFeB between the MgO barrier and W spacer is desireable for higher TMR.

Lastly, we did attempt to extend our multi-spacer design beyond two W spacers, but any higher number of spacers resulted in a steep degradation of film properties. It becomes more challenging to control the thickness of even thinner W layers and to optimize more Mg sacrificial layers. The extreme end of the multi-spacer approach is a CoFeBW alloy free layer, though this is likely to result in lower $M_S$ and lower Curie temperature.

\section{Conclusions}
In summary, by minimizing the W spacer thickness in the FL, we have eliminated the dead layer, thereby maximizing $M_S$. We obtained optimal magnetic properties by splitting the W spacer layer into two thinner layers. For the FL thickness ladder in this study, from 1.2~nm to 1.7~nm, highest $M_S$ was obtained for the thinnest FLs, while maintaining low damping constant and low inhomogenous broadening of the FMR linewidth, signifying a more uniform and better quality film.

From device measurements, $A_{\mathrm{ex}}$ was highest for the thinnest FLs with split-W spacer, and thus high $\Delta_{\mathrm{dw}}$ was maintained. We observed a thickness dependence of $J_{c0}$: the 1.2~nm FL has $\sim$30\% lower $J_{c0}$ than the 1.7~nm FL. The thinnest FLs with split-W spacer have the lowest $J_{c0}$ and highest spin-transfer-torque efficiency.

\section*{Data Availability}
The data that supports the findings of this study are available within the article.

\begin{acknowledgments}
The authors acknowledge Michael Gribelyuk, Loc Vinh and Xiaoyu Xu at the Western Digital Materials Laboratory for providing the high-resolution TEM analysis. 
\end{acknowledgments}

\bibliography{masterbibfile}

\end{document}